\documentclass[aps, prd, showpacs, floatfix, superscriptaddress, twocolumn, nofootinbib, reprint, longbibliography]{revtex4-2}
\usepackage{multirow, amsmath, hyperref, graphicx, booktabs}
\usepackage[T1]{fontenc}
\usepackage[utf8]{inputenc}
\usepackage{lmodern}
\usepackage{amsfonts}
\usepackage{array}
\usepackage{amssymb}
\usepackage{float}
\usepackage{url}
\usepackage{mathtools}
\usepackage{textcomp}
\usepackage{parskip}
\usepackage[dvipsnames]{xcolor}
\usepackage[autostyle]{csquotes}
\usepackage{orcidlink}
\usepackage[final]{microtype}
\usepackage{subcaption}
\usepackage{ulem}
\usepackage{empheq}
\hypersetup{
colorlinks=true,
citecolor=blue,
linkcolor=blue,
urlcolor=blue,
allbordercolors={0 0 0},
pdfborderstyle={/S/U/W 1}
}
\newcounter{mnotecount}[section]
\newcommand{\mnotex}[1]
{\protect{\stepcounter{mnotecount}}$^{\mbox{\footnotesize
$
\bullet$\themnotecount}}$ \marginpar{
\raggedright\tiny\em
$\!\!\!\!\!\!\,\bullet$\themnotecount: #1} }

\newcommand{\SA}{\author{Soham Acharya}
                 \email[]{acharyasoham@iitgn.ac.in}}
\newcommand{\SR}{\author{Shuvayu Roy}
                 \email[]{shuvayu.roy@iitgn.ac.in}}
\newcommand{\SSr}{\author{Sudipta Sarkar}
                 \email[]{sudiptas@iitgn.ac.in}}
                 
\newcommand{\IITGn}{Indian Institute of Technology, Gandhinagar, Gujarat 382055, India}

\begin{document}

\title{Stochastic Evolution of Primordial Black Holes to near-extremality in EFTs of Gravity}

\SA
\affiliation{\IITGn} 

\SR
\affiliation{\IITGn} 

\SSr
\affiliation{\IITGn}

\begin{abstract}
    The search for dark matter candidates includes primordial black holes (PBHs) as possible constituents. Recent studies show that some PBHs can survive to the present epoch by gaining angular momentum through Hawking radiation of photons and becoming extremal before complete evaporation. While this provides a plausible model in a two-derivative theory of gravity, additional issues arise in EFT-corrected theories of gravity. In such theories, a rapidly spinning black hole can lead to extremely high tidal forces on a near-horizon observer, with possible observational consequences. In this work, by modeling Hawking radiation as a biased random walk within an EFT of gravity, we show that nearly the same fraction of PBHs survives as in GR. We argue that the resultant near-horizon tidal effects should be detectable in future gravitational-wave observables.
\end{abstract}

\maketitle

\section{Introduction}\label{Introduction}

Cold dark matter (CDM) has remained one of the essential ingredients in our current understanding of the universe. Among the various objects that can constitute a CDM, black holes have remained one of the most longstanding candidates. However, to be considered as viable dark matter candidates, they must have formed before Big Bang Nucleosynthesis (BBN) and persist till the present epoch \cite{Chapline:1975ojl,Carr:2021bzv}. These conditions can possibly be satisfied by Primordial Black Holes (PBHs) formed well before BBN from the collapse of large density fluctuations, bubble collisions, or cosmic string loops \cite{Carr:2020xqk, Carr:2021bzv, Hektor:2018rul,Green2014, Jedamzik:1999am,Jenkins:2020ctp}. While a tentative lower bound has been suggested on the mass of these PBHs \cite{Hektor:2018rul, Korwar:2023kpy}, several studies have also investigated mechanisms that could allow PBH with a very low mass to have survived to the present day \cite{MacGibbon:1987my,Barrow:1992hq}. 

Recently, in \cite{Taylor:2024fvf}, the authors have proposed a possible survival mechanism based on spin evolution during Hawking evaporation. In this picture, the PBH mass decreases monotonically due to evaporation, while the angular momentum undergoes a biased random walk driven by the spin of the emitted quanta. As the spin of the black hole increases, the Hawking temperature decreases, and for a significant fraction of PBHs, the evolution leads up to an extremal Kerr state. The authors report that almost $22\%$ of an ensemble of PBHs modelled as Schwarzschild BHs spin up to extremality. At or near extremality, the Hawking temperature approaches zero, effectively halting further evaporation and producing long-lived or stable extremal Kerr black holes that could serve as dark matter candidates.

While these scenarios are compelling within the semi-classical framework, at the later stages of evaporation, the black hole mass reduces to the order of a few Planck masses. At such scales, due to the extremely low length-scales and hence, high energy scales being probed, it is expected that Einstein's two-derivative theory of gravity (called ``GR'' henceforth) would not explain the system sufficiently, and higher-derivative corrections should come into effect. These higher-derivative effects can be treated within an effective field theory (EFT) approach, remaining agnostic to the UV-complete theory of gravity. Although these EFT corrections are expected to alter the behavior of the black holes only perturbatively, there can arise cases where they lead to observationally significant effects. Recent investigations \cite{Horowitz:2023xyl,Horowitz:2024dch} have shown that EFT corrections to an extremal Kerr black hole in asymptotically-flat four-dimensional spacetime can lead to divergent tidal forces near its horizon. In case the black hole spins close to extremality, the tidal forces, which scale as an inverse power of its temperature, can be significantly large, signaling a breakdown of the EFT.

Motivated by these findings, we investigate whether primordial black holes in an EFT-corrected theory of gravity can evolve, through Hawking emission of photons, from initially non-rotating states to configurations with near-extremal angular velocity. For a physically viable region in the parameter space of EFT-coefficients, we estimate the fraction of primordial black holes that can undergo such an evolution and assess its sensitivity to higher-curvature effects. Further, we demonstrate that for a range of different initial mass and initial spin configurations, a similar number of primordial black holes evolve to near-extremal spins, thus showing the robustness of the spin-evolution mechanism even in the presence of EFT-corrections. Keeping in mind the possibly diverging tidal effects discussed in \cite{Horowitz:2023xyl,Horowitz:2024dch}, we then argue that if such near-extremal spinning black hole solutions of gravitational EFT constitute even a small fraction of the dark matter abundance, they would generically produce strong and potentially observable signatures, driven by the rapidly growing tidal forces in the near-horizon region. These effects would then offer a novel probe of both primordial black hole dark matter and the breakdown of effective field theory in strong-gravity regimes, which are relevant to the early-universe scenario. 

The paper is organized as follows. Sec. \ref{setup} discusses the EFT-corrected theory of gravity that we use to model the black holes, and the probability function used for the spin-evolution mechanism. We present the results of the simulations in Sec. \ref{Results} and conclude with a discussion on their possible consequences in Sec. \ref{Discussions}. Appendix \ref{AppA} lists the expressions of the EFT corrections to temperature, and Appendix \ref{AppB} elaborates upon the inclusion of EFT corrections to the probability function.

\section{Setup}\label{setup}
In a four-dimensional asymptotically flat spacetime, the Einstein-Hilbert Lagrangian can be corrected up to eight derivatives as \cite{Endlich:2017tqa,Reall:2019sah,Horowitz:2023xyl}
\begin{equation}\label{EftExP}
    \mathcal{L} = \frac{1}{16 \pi \kappa^2} (R + \eta \kappa^4 \mathcal{R}^3+\lambda \kappa^6 \mathcal{C}^2+\tilde{\lambda} \kappa^6 \tilde{\mathcal{C}}^2)
\end{equation}
with $\mathcal{R}^3 \equiv R^{ab}{}_{cd}\,R^{cd}{}_{ef}\,R^{ef}{}_{ab}$, $\mathcal{C} \equiv R_{abcd}R^{abcd}$, $\tilde{\mathcal{C}} \equiv R^{abcd}\,\epsilon_{ab}{}^{pq}\,R_{pqcd}$
and $\kappa = M_{\text{Pl}}^{-1}$, with $M_{\text{Pl}}$ denoting the Planck mass. Following \cite{Taylor:2024fvf}, we work in units where mass is measured in units of $M_{\text{Pl}}$. Following \cite{Reall:2019sah,Cano:2019ore}, the EFT corrected form of the Hawking temperature of a Kerr BH can be expressed in terms of the dimensionless spin parameter $\chi = J/\kappa^2M^2$ as
\begin{align}
T_{\mathrm{EFT}}
&= T_{\mathrm{GR}}
\Bigg[
1
+ \frac{\eta}{\kappa^4 M^4}\,\Delta T_{\eta}(\chi)
+ \frac{\lambda}{\kappa^6 M^6}\,\Delta T_{\lambda}(\chi)
\nonumber\\
&\hspace{2.5em}
+ \frac{\tilde{\lambda}}{\kappa^6 M^6}\,
  \Delta T_{\tilde{\lambda}}(\chi)
\Bigg],
\\
T_{\mathrm{GR}}(\chi)
&= \frac{1}{4\pi\kappa^2 M}
   \frac{\sqrt{1-\chi^2}}{1+\sqrt{1-\chi^2}}~.
\end{align}
The explicit forms of the EFT corrections to the temperature ($\Delta T_{\eta}(\chi),\Delta T_{\lambda}(\chi), \Delta T_{\tilde \lambda}(\chi)$) are mentioned in Appendix \ref{Sec:intro}. 
In GR, the extremal limit is marked by $\chi = \pm 1$, where the Hawking temperature vanishes $(T_{GR}=0)$. In the presence of EFT corrections, this extremal limit itself gets corrected to $\chi_{\text{EFT-ext}}^2 = 1 + \alpha\,\varepsilon + \mathcal{O}(\varepsilon^2)$, where $\varepsilon$ denotes the strength of the EFT coupling. However, as discussed in \cite{Reall:2019sah}, thermodynamic quantities such as the entropy and temperature become non-analytic in the EFT expansion at extremality. The perturbative EFT framework relies on a Taylor expansion in higher-derivative couplings, which is well-defined only for sub-extremal black holes $|\chi|<1$, where the temperature remains small but non-vanishing. Therefore, to maintain numerical stability and to ensure the validity of the perturbative expansion, we restrict our analysis to spins below a cut-off value $\chi_{\text{max}} = 0.99$.

To ensure that the perturbative hierarchy of the EFT corrections remains well-maintained within the range $0 \leq \chi \leq 1$, we impose a "budget-splitting" criterion on the fractional temperature corrections.

The overall signs of the EFT-coefficients $(\lambda >0, \tilde \lambda>0)$ are fixed independent of $\chi$ from physical arguments like causality of graviton propagation \cite{Gruzinov:2006ie} and the unitarity and analyticity of scattering amplitudes \cite{Bellazzini:2015cra}, whereas for $\eta<0,$ one has to assume the standard model as argued in \cite{Horowitz:2023xyl, Goon:2016mil}. By scanning over several different weight combinations, we find the following range of the coefficients to form a reasonable parameter space for our analysis: $\eta  \in [-1.85\times10^{-5},\,0], \, \lambda \in [0,\,1.1\times10^{-7}], \tilde{\lambda} \in [0,\,1.7\times10^{-8}]. $

Another crucial input in the construction of this probability evolution framework is the dependence of the probability function on the spin of the photon emitted by the black hole. During Hawking radiation from a rotating black hole, photons are more likely to be emitted with spin aligned to the black hole’s angular momentum \cite{Hawking:1975vcx}. Consequently, each emission event tends to reduce rather than increase the black hole’s spin. Thus, this process can be modeled as a biased random walk problem, favoring the black hole's spin being reduced. To keep the model tractable, we consider only the dominant s-wave emission channel, which sufficiently captures the qualitative features of Hawking photon emission relevant to the stochastic spin evolution. Further, we restrict to the $m=\pm 1$ modes \cite{Page:1976df, Hawking:1975vcx} and expand all spin-dependent quantities to linear order in $\chi$. For a black hole with dimensionless spin $\chi$, the probability that its spin increases (decreases) after photon emission is parametrized by
\begin{align}\label{eqEFTProb1}
&P_{\substack{\uparrow \uparrow \\ \uparrow \downarrow}}(\chi) = \frac{1}{2} \mp\left(1+C_\text{EFT}\right) \chi .\\
&C_\text{EFT} = \frac{1}{2}\frac{\eta}{M^4\kappa^4}+\frac{13 }{44}\frac{\lambda}{M^6\kappa^6}-\frac{108}{11}\frac{\tilde \lambda}{M^6\kappa^6}
\end{align}
All the EFT corrections in this function $P_{\substack{\uparrow \uparrow \\ \uparrow \downarrow}}(\chi)$ enter through the temperature, angular velocity, and the greybody factor. While the EFT corrections to the first two quantities are known explicitly, the precise dependence of the greybody factor on the black hole spin $\chi$ and radiation frequency remains unavailable. Nevertheless, integrating over the full frequency range and guided by the low-spin analysis of \cite{Nomura:2012cx}, it is reasonable to expect the EFT effects to enter the greybody factor in an effective, parametrized form as adopted here. Moreover, parametrizing the EFT corrections at spins following \cite{Taylor:2024fvf}, and using the boundary conditions $P\substack{\uparrow\uparrow}(\chi = 1) = 0, \, P\substack{\uparrow\uparrow}(\chi = -1) = 1 $, the probability function can be modeled as
\begin{equation}\label{eqEFTProb2}
P_{\substack{\uparrow \uparrow \\ \uparrow \downarrow}}(\chi) = \frac{1}{2}\mp\left(1+C_\text{EFT}\right)\chi
\pm\left(\frac{1}{2}+C_\text{EFT}\right)\chi\,|\chi|~.
\end{equation}
In the presence of EFT corrections, the extremal spin generically shifts away from its GR value, so that $\chi_{\rm ext}$ no longer coincides with $\chi=1$. Hence, in principle, the appropriate boundary condition to fix the form of $P_{\uparrow\uparrow,\uparrow\downarrow}$ should take $\chi_{\rm ext}$ into account. However, we note that in the absence of an explicit expression for the EFT-corrected extremality bound $\chi_{\rm ext}$ in literature, and since we are conducting a qualitative study restricted to black holes reaching spins below extremality, we have used the boundary condition at $\chi=1$ as a proxy. 

With the probability model of photon emission that changes angular momentum in place and the specified EFT parameter space, we evolve an ensemble of primordial black holes under Hawking radiation. The evolution extends the stochastic framework of \cite{Taylor:2024fvf} to incorporate EFT-corrected thermodynamics. We initialize a large ensemble of EFT-corrected Schwarzschild black holes with mass $M_0$ and zero angular momentum ($J_0 = 0$), allowing each to undergo a sequence of discrete emission events that simultaneously decrease its mass and alter its angular momentum. The emitted energy quanta are sampled from a Planck distribution with each emission step corresponding to a mass loss $\delta M = x.T_{\rm{EFT}}$, and a change of angular momentum from $J \to J \pm 1$.
Here $x$ is a dimensionless ratio given by the energy of the emitted quanta per unit temperature. The evolution proceeds iteratively, updating both $M$ and $J$ and terminates when either the black hole reaches the imposed sub-extremal cut-off ($\chi = \chi_{\max} = 0.99$) or its mass drops below the cut-off mass $M_{\mathrm{cut}}$. For each simulation, we record the final mass, spin, and temperature, and compute the fraction of black holes that reach the near-extremal regime to study how this fraction depends on the EFT parameters.

Following the biased random walk prescription of Eq. \ref{eqEFTProb2}, and a Planckian spectrum for the energies of the emitted quanta, we initialize an ensemble of $10^6$ trajectories, evolving them from an initial state of mass $M_0 = 10 M_\text{Pl}$ and angular momentum $J_0 = 0$. Similar to \cite{Taylor:2024fvf}, we performed the same analysis for a range of initial masses while keeping the initial spin fixed at zero, as well as for a range of initial spins while keeping the initial mass fixed. In both cases, we found that the resulting spin-up probabilities remain nearly identical, as shown in Sec. \ref{Results}.

\section{Results}\label{Results}

\begin{figure*}
    \centering
    \begin{subfigure}{0.45\textwidth}
        \centering
        \includegraphics[width=\textwidth]{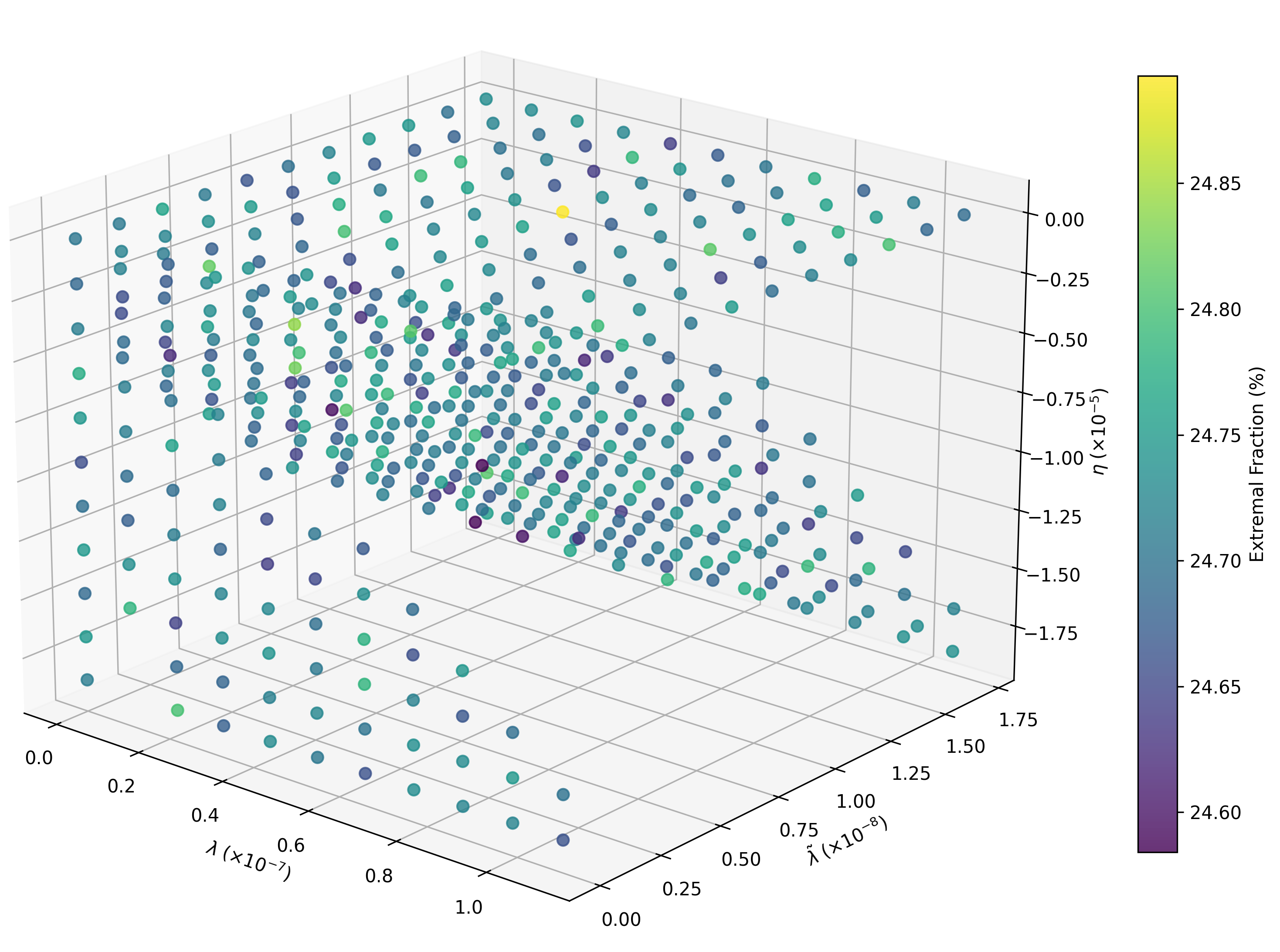}
        \caption{Azimuthal angle = $-50^{\circ}$.}
        \label{fig:extremal_fraction_1}
    \end{subfigure}
    \hfill 
    \begin{subfigure}{0.45\textwidth}
        \centering
        \includegraphics[width=\textwidth]{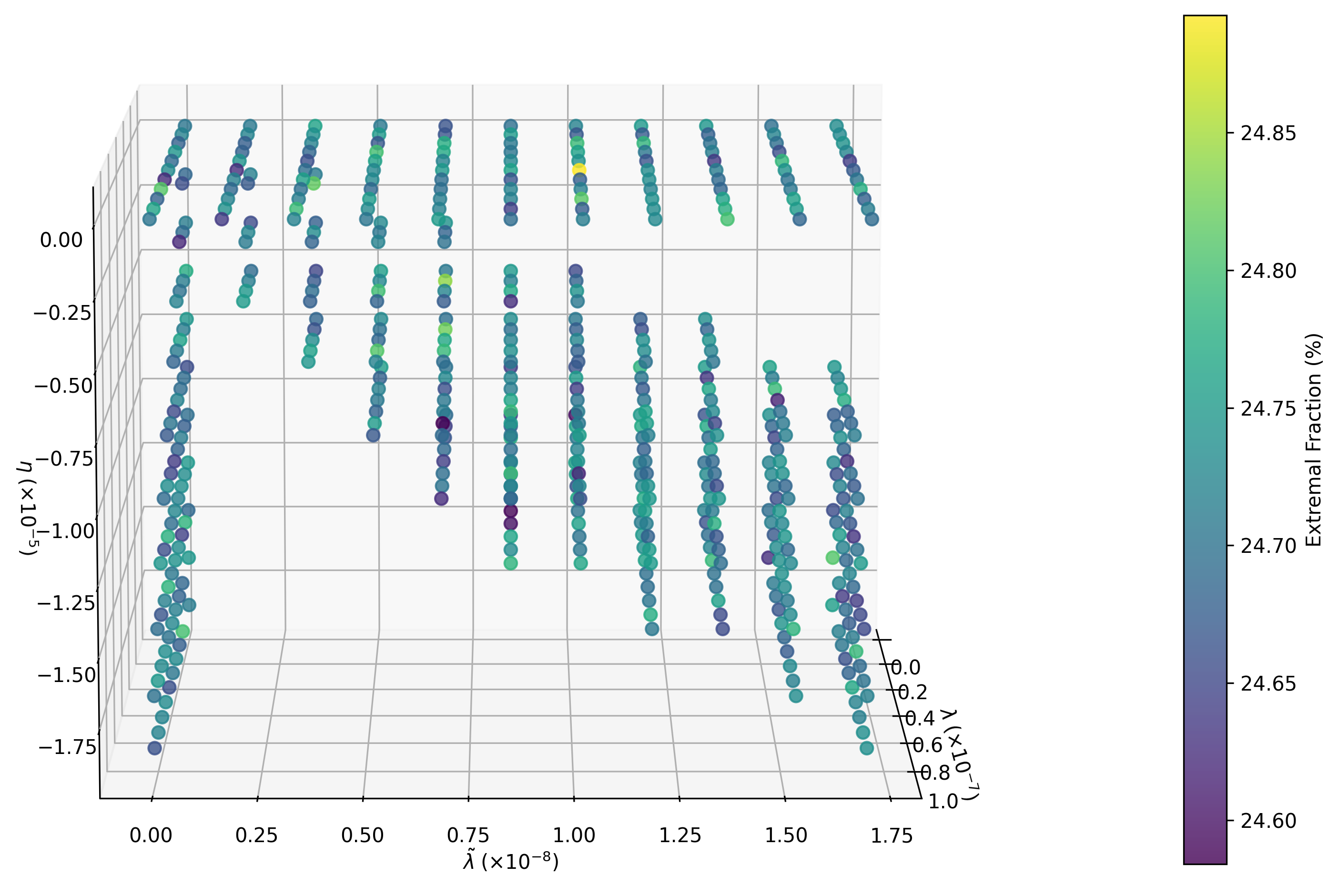}
        \caption{Azimuthal angle = $0^{\circ}$.}
        \label{fig:extremal_fraction_2}
    \end{subfigure}
    \caption{Fraction $(\%)$ of near-extremal PBHs across the EFT parameter space from two different azimuthal angles.}
    \label{fig:extremal_fraction}
\end{figure*}

The parameter space spanned by $(\eta,\lambda,\tilde{\lambda})$ is explored by discretizing each parameter into $11$ uniformly spaced values, yielding a total of $11^3=1331$ points across the full parameter space. However, for a subset of these points, the values of $\eta$ and $\lambda$ are such that, for black holes at the final stage of the stochastic evolution ($M\simeq1.3\,M_{\rm Pl}$), the metric perturbation sourced by the $\mathcal{O}(\eta^2)$ term in the effective Lagrangian can become larger than the contribution from the $\mathcal{O}(\lambda)$ term. In these cases, the linearized EFT expansion in \eqref{EftExP} loses validity and therefore breaks down \cite{Horowitz:2022mly}.
Similarly, there exist regions of parameter space in which the hierarchical structure of the temperature expansion is violated, with contributions from the eighth-derivative operators dominating over those from the sixth-derivative operators. Such parametric combinations are similarly excluded from our analysis. These consistency requirements account for the gaps observed in the parameter-space distribution shown in Fig.~\ref{fig:extremal_fraction}. We are therefore left with approximately $500$ viable points in parameter space where the stochastic evolution proceeds within the regime of EFT validity. 

As a reference point for quantifying the impact of EFT corrections, we first evaluate the fraction of trajectories reaching $\chi_\text{max}$ in the baseline case $\eta=\lambda=\tilde\lambda=0$. Ref. \cite{Taylor:2024fvf} found approximately $22\%$ of black holes evolving to true extremality ($\chi = 1$), whereas, in our setup, we observe about $24.7\%$ reach $\chi_\text{max} = 0.99$, hereafter referred to as near-extremal black holes (NEBHs). This enhanced fraction is expected, as the introduction of an effective cutoff replaces the strict extremality condition, permitting a larger set of trajectories to meet the near-extremal criterion before the evolution is halted. We expect this fraction to decrease toward the semiclassical value if the evolution were extended all the way to $\chi_{\rm ext}$.

\begin{figure}
    \centering
    \vspace{0.5cm}  
    \includegraphics[width=0.9\linewidth]{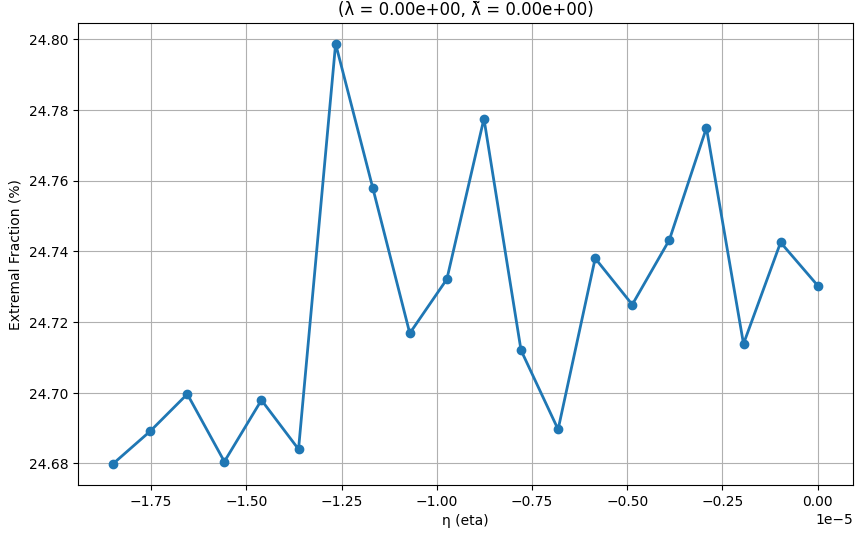}
    \caption{Fraction $(\%)$ of near-extremal PBHs as a function of $\eta$ for fixed $\lambda = 0$ and $\tilde{\lambda} = 0$.}
    \label{fig2}
\end{figure}

In the presence of EFT corrections, the fraction of NEBHs lies within $24.6\%$–$24.9\%$. The distribution of these fractions across the sampled phase space is shown in Fig.~\ref{fig:extremal_fraction}. To provide a clearer visualization of the structure of the parameter space, we present two complementary perspectives of the same dataset corresponding to two different azimuthal viewing angles. In Fig. \ref{fig2}, we plot the fraction of PBHs that evolve to $\chi=\chi_\text{max}$ for a range of different $\eta$'s with $\lambda=\tilde \lambda=0$. 

\begin{figure*}[t]
    \centering
    \includegraphics[width=\textwidth]
    {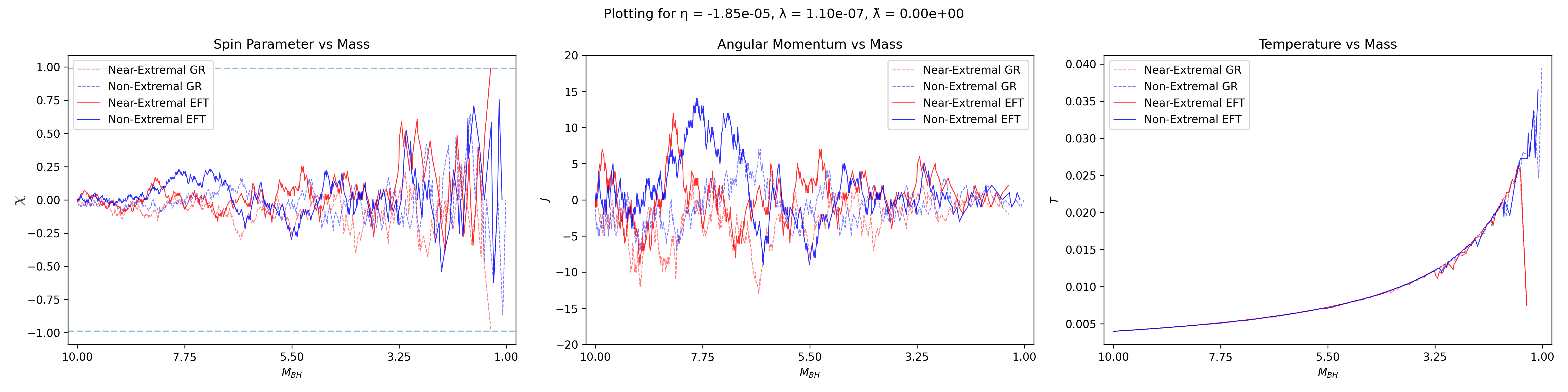}
    \caption{Evolution of $\chi$, $J$, and $T$ as functions of the black hole mass.}
    \label{fig:trajectories}
\end{figure*}

We next examine the evolution of representative individual trajectories, considering the following cases: (i) a PBH in GR evolving to near extremality; (ii) a PBH in GR evaporating completely before becoming near extremal; (iii) a PBH in the EFT corrected theory that evolves to near extremality; and (iv) a PBH in the EFT corrected theory that evaporates completely prior to reaching near extremality. In Fig.~\ref{fig:trajectories}, we plot the evolution of the dimensionless spin parameter $\chi$, the angular momentum $J$, and the black hole temperature $T$ versus mass during the evaporation process.

We find that the spin parameter $\chi$ remains close to zero for most of the black hole lifetime, indicating that PBHs remain effectively EFT-corrected Schwarzschild ($\chi^2 \ll 1$) until their masses decrease to within a few times the Planck mass. A rapid spin-up occurs only during the final stages of evaporation, driven by the emission of the last few Hawking quanta.

\begin{figure}[!h]
    \centering
    \includegraphics[width=\linewidth]{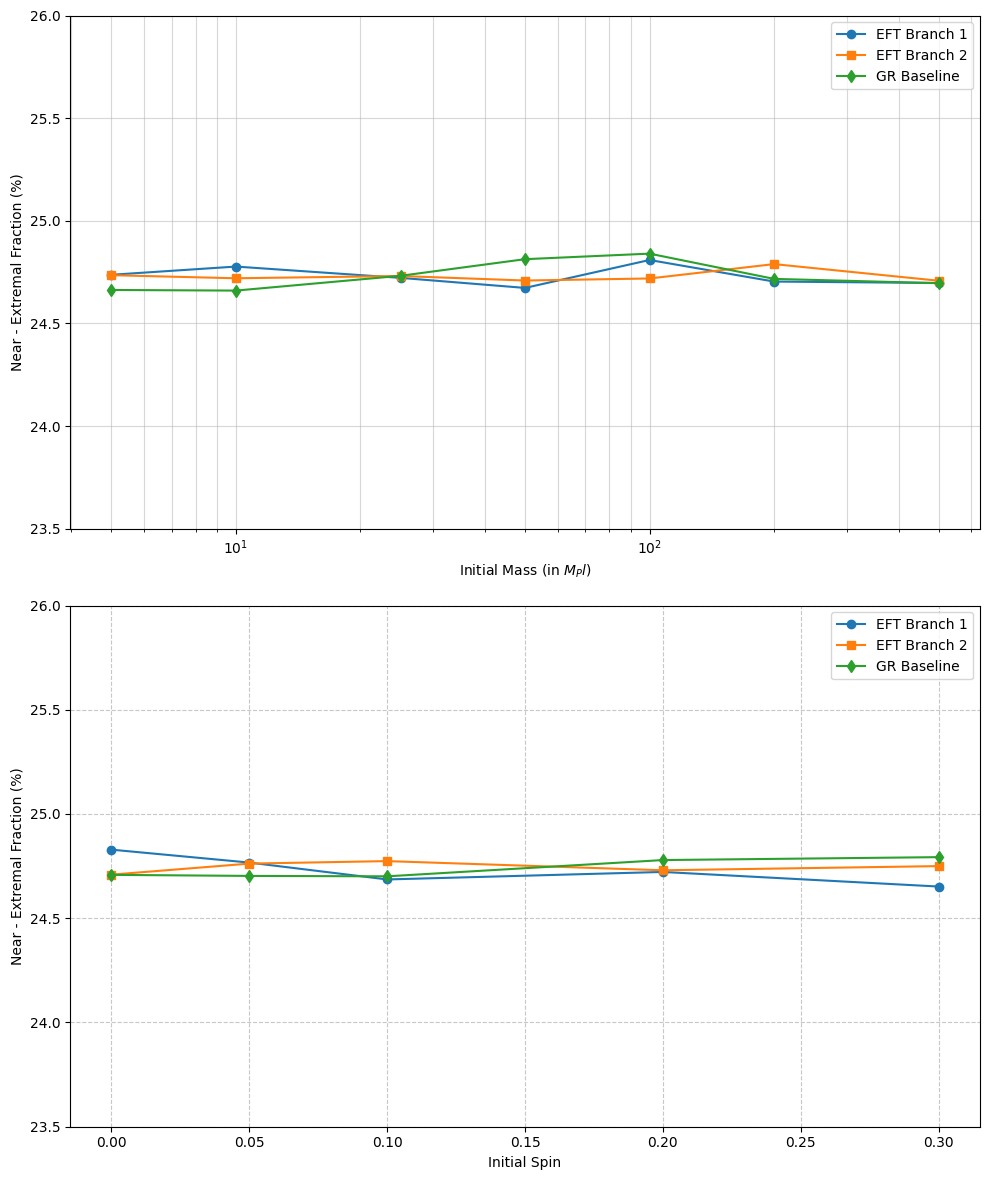}
    \caption{Percentage of black holes reaching the near-extremal limit when varying initial mass (at zero initial spin) and varying initial spin (at $10\,M_{\mathrm{Pl}}$). \newline \textbf{EFT Branch 1:} $\eta = -1.85 \times 10^{-5}$, $\lambda = 3.30 \times 10^{-8}$, $\tilde{\lambda} = 1.70 \times 10^{-8}$. \newline \textbf{EFT Branch 2:} $\eta = -1.66 \times 10^{-5}$, $\lambda = 1.10 \times 10^{-7}$, $\tilde{\lambda} = 1.70 \times 10^{-8}$.}
    \label{fig:ini_cond_Probab}
\end{figure}

We also find that the fraction of trajectories evolving toward the near-extremal regime remains largely insensitive to variations in the initial mass over the explored range, consistent with the qualitative statement in Sec. \ref{setup} earlier. A systematic scan of the full EFT parameter space over a broad mass range is computationally expensive due to longer evaporation timescales, making simulations for masses much larger than $10^3 M_{\mathrm{Pl}}$ prohibitively expensive within the present framework. The attached plot, therefore, serves as a representative demonstration that the reported near-extremal fraction remains stable against changes in the initial mass within the accessible computational range. Regarding the initial spin, our original motivation, following Taylor et al., was to study the spin evolution of small-mass primordial black holes modeled as Schwarzschild, for which the initial angular momentum is expected to be small. However, we have now additionally performed simulations with nonzero initial spins (for $M = 10\,M_{\mathrm{Pl}}$) and found that the near-extremal fraction remains essentially unchanged across the small-spin regime considered. (See Fig.\ref{fig:ini_cond_Probab})

These results demonstrate that the stochastic evolution mechanism permits a substantial fraction of PBHs to reach near extremal values of spin and temperature. As argued in \cite{Horowitz:2022mly,Horowitz:2023xyl,Horowitz:2024dch}, such near-extremal configurations are associated with extremely large tidal forces experienced by infalling observers near the horizon. If PBHs persist in the present epoch and contribute appreciably to the dark matter abundance, such enhanced tidal effects could imprint observable signatures in gravitational wave signals. This leads to two distinct possibilities: either the large tidal forces render rapidly rotating PBHs unstable before extremality is reached, disfavoring extremally spinning PBHs as viable dark matter candidates, or a population of long-lived, near-extremal PBHs exists, detectable by future gravitational wave observatories.

We also emphasize that our analysis is restricted to the domain of the couplings $\eta$, $\lambda$, and $\tilde{\lambda}$ where the EFT corrections remain consistent, satisfying all causality and unitarity constraints, thus ensuring the validity of the results established in Refs.~\cite{Horowitz:2022mly, Horowitz:2023xyl}.

\section{Summary and Outlook}\label{Discussions}

In this work, we assess the viability of extremal rotating black holes as dark matter candidates when effective field theory corrections to gravity become important. Building on the semiclassical treatment of Hawking radiation in Ref. \cite{Taylor:2024fvf}, we perform a population-level analysis showing that in EFT-corrected spacetimes, a significant fraction of black holes can evolve from initially non-rotating or moderately rotating states to near extremality through a biased stochastic spin up driven by Hawking emission. This provides a concrete dynamical mechanism by which primordial black holes with EFT corrections can acquire near extremal spins and potentially contribute to the dark matter abundance.

Once higher derivative EFT corrections to the gravitational action are included, Ref. \cite{Horowitz:2023xyl} shows that for certain allowed signs of the EFT coefficients, extremal rotating black holes in asymptotically flat spacetime develop curvature singularities on the horizon. Even for slightly sub-extremal configurations, the spacetime can exhibit large near-horizon tidal forces despite finite curvature invariants. A rough estimate of the magnitudes of these divergences can be made as follows. Following Refs.~\cite{Horowitz:2022mly,Horowitz:2023xyl,Horowitz:2024dch}, the dominant near-horizon tidal components of the Weyl tensor scale as
 \begin{equation}\label{Weyl_Temperature}
 C_{\rho A \rho B} \sim T^{\gamma-2},
 \end{equation}
 where $\rho$ denotes a near-horizon radial coordinate, $A,B$ label angular directions, $T$ is the Hawking temperature, and $\gamma$ encodes the near-horizon scaling behavior.
This motivates defining a dimensionless measure of EFT-induced tidal enhancement by comparing the EFT-corrected Kerr geometry to its GR counterpart,
 \begin{equation}
 \delta C = \frac{C^{\text{EFT}}_{\rho A \rho B}}{C^{\text{GR}}_{\rho A \rho B}}
 \equiv K_1 \frac{{T_{\rm EFT}}^{\gamma_{EFT}-2}}{{T_{\rm GR}}^{\gamma_{GR}-2}},
 \end{equation}
where $K_1$ is an $\mathcal{O}(1)$ coefficient. 
To evaluate this ratio, we consider a representative near-extremal configuration extracted from our simulations, with dimensionless spin $\chi = 0.99$ and mass $M \simeq 1.3,M_{\rm Pl}$, corresponding to the typical mass of EFT-corrected near-extremal Kerr black holes at the endpoint of the stochastic evolution. \footnote{The values of the scaling exponent are taken from Ref.~\cite{Horowitz:2023xyl}, where it was shown that $\gamma_{EFT}<2$ for extremal black holes in the presence of higher-derivative EFT corrections. Although these values strictly apply at extremality, we adopt them here as a rough proxy to estimate the magnitude of tidal effects in the near-extremal regime. As we are interested only in an order-of-magnitude estimate of the tidal enhancement, this approximation is sufficient despite the extrapolation away from exact extremality.}
We evaluate the tidal-force enhancement for a small set of representative EFT couplings drawn from the parameter space explored in our simulations. For example, for a few coupling choices of $(\eta,\lambda,\tilde \lambda)$, viz. 
$(-1.48\times10^{-5},\,9.9\times10^{-8},\,1.53\times10^{-8})$, $(-1.66\times10{-5},9.9\times10^{-8},1.7\times{-8})$, $(-1.85\times10^{-5},\,1.1\times10^{-7},\,1.7\times10^{-8})$, 
we find $\delta C$ of order $10$, with values ranging between approximately $9$ and $12$. We find
\begin{equation}
\delta C \sim \mathcal{O}(10^1),
\end{equation}
which implies that even in the sub-extremal regime ($\chi\simeq0.99$), the tidal forces in the EFT-corrected geometry can be enhanced by approximately an order of magnitude relative to Kerr. Importantly, since the exponent $\gamma_{\rm EFT}-2$ is negative, this ratio scales inversely with temperature. As a result, as the black hole approaches closer to extremality and the temperature decreases further, the corresponding measure of tidal forces grows rapidly and could diverge in the extremal limit. 
While we were unable to simulate the evolution up to the exact extremal limit due to the limitations of the framework, the structure of the Weyl tensor's components in \eqref{Weyl_Temperature} and the aforementioned estimates indicate a possible divergence at the extremal limit.

These results and estimates lead to interesting and physically significant consequences. While EFT-corrected dynamics allow the formation of a sizable population of near-extremal black holes through Hawking radiation-driven spin-up, the resulting objects could be accompanied by strong near-horizon tidal effects that may lead to distinctive, potentially observable signatures beyond general relativity. One possible method to further understand the impact of these divergent tidal forces would be to calculate the modified Hill radius of a secondary body orbiting such black holes \cite{Hills1975, Xin2026}. This requires computing the tidal tensor by projecting the Weyl tensor onto a local orthonormal tetrad along the orbit, which in turn necessitates obtaining the complete modified black-hole geometry. Such a calculation can be carried out using pseudo-spectral methods and is part of ongoing work. Investigating the resulting tidal stripping of matter bound to the secondary could further lead to distinctive dephasing signatures in gravitational-wave signals.

This work opens several directions for future study. In particular, following Ref. \cite{Barausse:2018vdb}, it would also be interesting to investigate whether these instabilities source a stochastic gravitational wave background detectable by current or future experiments. Such observations could provide an independent probe of the EFT parameters $\eta$, $\lambda$, and $\tilde{\lambda}$ and further constrain scenarios in which extremal or near-extremal spinning black holes contribute to dark matter.

\section*{Acknowledgments} 

We acknowledge the use of the \textit{NOETHER} workstation at the Department of Physics, IIT Gandhinagar. The research of SS is supported by the Department of Science and Technology, Government of India, under the ANRF CRG Grant (No. CRG/2023/000545).

\appendix

\begin{widetext}
\section{EFT Corrections to temperature}\label{AppA}

\label{Sec:intro}

The EFT corrected Hawking temperature can be written as \cite{Reall:2019sah}: 

\begin{equation}
\begin{split}
      T_{\mathrm{EFT}}
  = &T_{\mathrm{Kerr}}\!\left(
    1 +\frac{\eta}{\kappa^4 M^4}\,\Delta T_{\eta}\right.\\
      &\left.+ \frac{\lambda}{\kappa^6 M^6}\,\Delta T_{\lambda}
      + \frac{\tilde{\lambda}}{\kappa^6 M^6}\,\Delta T_{\tilde{\lambda}}
  \right)    
\end{split}
\end{equation}
\begin{subequations}\label{Temp_EFT_Coeff}
\renewcommand{\theequation}{\theparentequation\alph{equation}}
\begin{equation}
\Delta T_\eta(\chi)=
\frac{1}{7}\!\left(
\frac{
  5 - 22\chi^2 + 16\chi^4
  + 2\sqrt{1-\chi^2}\,\bigl(1 - 6\chi^2 + 4\chi^4\bigr)
}{
  (1-\chi^2)\,\bigl(1+\sqrt{1-\chi^2}\bigr)^3
}\right).
\end{equation}
\begin{equation}
\begin{aligned}
\Delta T_{\lambda}(\chi)
&= -\frac{2}{5}\!\left(
\frac{\displaystyle
\begin{aligned}
&40950 - 71925\chi^2 + 31290\chi^4 - 1125\chi^6 + 370\chi^8 - 520\chi^{10} + 480\chi^{12} - 128\chi^{14} \\
&\quad - \sqrt{1-\chi^2}\,\bigl(40950 - 62475\chi^2 + 23415\chi^4 - 810\chi^6 + 280\chi^8 - 400\chi^{10} + 256\chi^{14}\bigr) \\
&\quad + 1575(1-\chi^2)\,\Bigl[26 - 31\chi^2 + 8\chi^4 - \sqrt{1-\chi^2}\,\bigl(26 - 11\chi^2\bigr)\Bigr]\frac{\arcsin\chi}{\chi}
\end{aligned}
}{
\displaystyle
\chi^{10}(1-\chi^2)\,\bigl(\sqrt{1+\chi}+\sqrt{1-\chi}\bigr)^6
}\right).
\end{aligned}
\end{equation}
\begin{equation}
\begin{aligned}
\Delta T_{\tilde{\lambda}}(\chi)
&= -\frac{8}{5}\!\left(
\frac{\displaystyle
\begin{aligned}
&40950 - 71925\chi^2 + 31290\chi^4 - 1125\chi^6 + 50\chi^8 + 520\chi^{10} - 480\chi^{12} + 128\chi^{14} \\
&\quad - \sqrt{1-\chi^2}\,\bigl(40950 - 62475\chi^2 + 23415\chi^4 - 810\chi^6 - 40\chi^8 + 400\chi^{10} - 256\chi^{14}\bigr) \\
&\quad + 1575(1-\chi^2)\,\Bigl[26 - 31\chi^2 + 8\chi^4 - \sqrt{1-\chi^2}\,\bigl(26 - 11\chi^2\bigr)\Bigr]\frac{\arcsin\chi}{\chi}
\end{aligned}
}{
\displaystyle
\chi^{10}(1-\chi^2)\,\bigl(\sqrt{1+\chi}+\sqrt{1-\chi}\bigr)^6
}\right).
\end{aligned}
\end{equation}
\end{subequations}
\end{widetext}

Although each correction function in \eqref{Temp_EFT_Coeff} appears to diverge as $\chi \to 0$ due to explicit $\chi^{-n}$ factors, these divergences cancel upon expansion, yielding a smooth Schwarzschild limit. The resulting temperature correctly reproduces the EFT-corrected Schwarzschild result. Moreover, the odd-parity six-derivative term proportional to $\tilde{\lambda}$ gives no contribution in this limit, since its associated parity-odd curvature invariant $R_{abcd}\tilde{R}^{abcd}$ vanishes for a spherically symmetric background. 

\section{EFT Corrections to biasing probability}\label{AppB}

A central input to this stochastic evolution framework is the probability for the emission of a quantum that carries angular momentum away from the black hole. Specifically, our objective is to construct a continuous, monotonic probability density function that dictates the spin-down bias, i.e., the statistical preference for the black hole to emit quanta with angular momentum anti-aligned with its own spin. This reduces the spin evolution to a biased random walk problem.\\

To model this bias in a tractable way, we adopt a set of simplifying assumptions similar to those in \cite{Taylor:2024fvf}, restricting our focus to the dominant emission channel for photons. Because we are explicitly tracking the evolution of the black hole's spin, we isolate the specific modes that carry axial angular momentum and therefore dictate the spin-down bias. Modes with $m=0$ carry no axial angular momentum \cite{Page:1976df} and do not contribute to the bias; therefore, we begin by calculating the emission probability derived from the Hawking number flux strictly for photons in the $m=\pm 1$ modes.
The number flux of photons emitted in a specific mode is given by:
\begin{equation}
\langle N_{lm} \rangle = \frac{\Gamma_{lm}(\omega, \chi)}{\exp\left(\frac{\omega - m \Omega_H}{T}\right) - 1}
\end{equation}
where $\Gamma_{lm}(\omega, \chi)$ is the spin and frequency-dependent greybody factor, $T$ is the black hole temperature, and $\Omega_H$ is the angular velocity of the horizon. From this, the probability that a given emitted photon is anti-aligned ($P_{\uparrow\downarrow}$) or aligned ($P_{\uparrow\uparrow}$) with the black hole's spin is constructed as the ratio of that specific mode's number flux to the total number flux in the angular-momentum-carrying channels. 

To model the leading-order bias, we utilize the linearly expanded forms for the EFT-corrected temperature and angular velocity derived from \cite{Reall:2019sah}. By expanding the resulting flux equations to linear order in the dimensionless spin parameter $\chi$, the frequency dependence and the exact functional form of the greybody factors can be factorized out, as all such dependencies are absorbed into a single undetermined constant, $C_0$. This yields the linear expanded form for the emission probability:

\begin{align}\label{eqEFTProb1}
&P_{\substack{\uparrow \uparrow \\ \uparrow \downarrow}}(\chi) = \frac{1}{2} \mp C_0 \left(1+C_\text{EFT}\right) \chi .\\
&C_\text{EFT} = \frac{1}{2}\frac{\eta}{M^4\kappa^4}+\frac{13 }{44}\frac{\lambda}{M^6\kappa^6}-\frac{108}{11}\frac{\tilde \lambda}{M^6\kappa^6}
\end{align}

A direct expansion beyond the linear order reveals that the next non-vanishing spin-dependent correction in the general relativistic limit appears only at the cubic order, $\mathcal{O}(\chi^3)$. While these cubic terms generically introduce required curvature into the probability profile as the spin approaches extremality, truncating the expansion at $\mathcal{O}(\chi^3)$ typically spoils the monotonicity of the probability over the full physical domain $\chi \in [-1, 1]$. A non-monotonic probability distribution would violate the strict physical boundary conditions required for the emission states: $P_{\uparrow\uparrow}(1) = 0$ (emission of aligned photons is forbidden at extremality) and $P_{\uparrow\uparrow}(-1) = 1$.

Therefore, we seek a minimal non-linear completion that captures the required curvature while preserving both the monotonicity and the asymptotic conditions of the probability. Following the authors in \cite{Taylor:2024fvf}, we introduce a corrective term by hand. This is achieved by adding a term of the form $C_1 \chi |\chi|$ to the linear contribution:  
\begin{equation}
    P_{\uparrow\downarrow}(\chi) = \frac{1}{2} \mp C_0 (1 + C_{EFT}) \chi \pm C_1 \chi |\chi|
\end{equation}

Applying the boundary conditions allows us to completely determine the coefficient $C_1$:  

\begin{equation}
    C_1 = -\frac{1}{2} - C_0 - C_0 C_{EFT}
\end{equation}

As seen from this relation, $C_0$ remains a free parameter parametrizing the baseline bias. We want to choose $C_0$ such that, when the higher-derivative EFT parameters ($\eta, \lambda, \tilde{\lambda}$) are turned off, $P_{\uparrow\downarrow}(\chi)$ smoothly reduces to the standard GR probability model quoted by Taylor et al. \cite{Taylor:2024fvf}. This leads us to $C_0 = -1$, which yields our final probability expression used for the simulations:  $$P_{\uparrow\downarrow}(\chi) = \frac{1}{2} \mp (1 + C_{EFT}) \chi \pm \left( \frac{1}{2} + C_{EFT} \right) \chi |\chi|$$.

\bibliography{ref}

@article{Carr:2021bzv,
    author = "Carr, Bernard and Kuhnel, Florian",
    title = "{Primordial black holes as dark matter candidates}",
    eprint = "2110.02821",
    archivePrefix = "arXiv",
    primaryClass = "astro-ph.CO",
    doi = "10.21468/SciPostPhysLectNotes.48",
    journal = "SciPost Phys. Lect. Notes",
    volume = "48",
    pages = "1",
    year = "2022"
}

@article{Carr:2020xqk,
    author = "Carr, Bernard and Kuhnel, Florian",
    title = "{Primordial Black Holes as Dark Matter: Recent Developments}",
    eprint = "2006.02838",
    archivePrefix = "arXiv",
    primaryClass = "astro-ph.CO",
    doi = "10.1146/annurev-nucl-050520-125911",
    journal = "Ann. Rev. Nucl. Part. Sci.",
    volume = "70",
    pages = "355--394",
    year = "2020"
}

@article{Hektor:2018rul,
    author = {Hektor, Andi and H{\"u}tsi, Gert and Raidal, Martti},
    title = "{Constraints on primordial black hole dark matter from Galactic center X-ray observations}",
    eprint = "1805.06513",
    archivePrefix = "arXiv",
    primaryClass = "astro-ph.CO",
    doi = "10.1051/0004-6361/201833483",
    journal = "Astron. Astrophys.",
    volume = "618",
    pages = "A139",
    year = "2018"
}

@article{Barrow:1992hq,
    author = "Barrow, John D. and Copeland, Edmund J. and Liddle, Andrew R.",
    title = "{The Cosmology of black hole relics}",
    reportNumber = "NSF-ITP-92-24, SUSSEX-AST-92-3-1",
    doi = "10.1103/PhysRevD.46.645",
    journal = "Phys. Rev. D",
    volume = "46",
    pages = "645--657",
    year = "1992"
}

@inbook{Green2014,
  title = {Primordial Black Holes: Sirens of the Early Universe},
  ISBN = {9783319108520},
  ISSN = {2365-6425},
  url = {http://dx.doi.org/10.1007/978-3-319-10852-0_5},
  DOI = {10.1007/978-3-319-10852-0_5},
  booktitle = {Quantum Aspects of Black Holes},
  publisher = {Springer International Publishing},
  author = {Green,  Anne M.},
  year = {2014},
  month = nov,
  pages = {129–149}
}

@article{Jedamzik:1999am,
    author = "Jedamzik, K. and Niemeyer, Jens C.",
    title = "{Primordial black hole formation during first order phase transitions}",
    eprint = "astro-ph/9901293",
    archivePrefix = "arXiv",
    doi = "10.1103/PhysRevD.59.124014",
    journal = "Phys. Rev. D",
    volume = "59",
    pages = "124014",
    year = "1999"
}

@article{Jenkins:2020ctp,
    author = "Jenkins, Alexander C. and Sakellariadou, Mairi",
    title = "{Primordial black holes from cusp collapse on cosmic strings}",
    eprint = "2006.16249",
    archivePrefix = "arXiv",
    primaryClass = "astro-ph.CO",
    reportnumber = "KCL-PH-TH/2020-30",
    month = "6",
    year = "2020",
    journal = ""
}

@article{Korwar:2023kpy,
    author = "Korwar, Mrunal and Profumo, Stefano",
    title = "{Updated constraints on primordial black hole evaporation}",
    eprint = "2302.04408",
    archivePrefix = "arXiv",
    primaryClass = "hep-ph",
    doi = "10.1088/1475-7516/2023/05/054",
    journal = "JCAP",
    volume = "05",
    pages = "054",
    year = "2023"
}

@article{Chapline:1975ojl,
    author = "Chapline, George F.",
    title = "{Cosmological effects of primordial black holes}",
    doi = "10.1038/253251a0",
    journal = "Nature",
    volume = "253",
    number = "5489",
    pages = "251--252",
    year = "1975"
}

@article{Cano:2019ore,
    author = "Cano, Pablo A. and Ruip{\'e}rez, Alejandro",
    title = "{Leading higher-derivative corrections to Kerr geometry}",
    eprint = "1901.01315",
    archivePrefix = "arXiv",
    primaryClass = "gr-qc",
    reportNumber = "IFT-UAM/CSIC-19-2",
    doi = "10.1007/JHEP05(2019)189",
    journal = "JHEP",
    volume = "05",
    pages = "189",
    year = "2019",
    note = "[Erratum: JHEP 03, 187 (2020)]"
}

@article{Taylor:2024fvf,
    author = "Taylor, Quinn and Starkman, Glenn D. and Hinczewski, Michael and Mihaylov, Deyan P. and Silk, Joseph and de Freitas Pacheco, Jose",
    title = "{Extremal Kerr black hole dark matter from Hawking evaporation}",
    eprint = "2403.04054",
    archivePrefix = "arXiv",
    primaryClass = "gr-qc",
    doi = "10.1103/PhysRevD.109.104066",
    journal = "Phys. Rev. D",
    volume = "109",
    number = "10",
    pages = "104066",
    year = "2024"
}

@article{Horowitz:2023xyl,
    author = "Horowitz, Gary T. and Kolanowski, Maciej and Remmen, Grant N. and Santos, Jorge E.",
    title = "{Extremal Kerr Black Holes as Amplifiers of New Physics}",
    eprint = "2303.07358",
    archivePrefix = "arXiv",
    primaryClass = "hep-th",
    doi = "10.1103/PhysRevLett.131.091402",
    journal = "Phys. Rev. Lett.",
    volume = "131",
    number = "9",
    pages = "091402",
    year = "2023"
}

@article{Horowitz:2024dch,
    author = "Horowitz, Gary T. and Kolanowski, Maciej and Remmen, Grant N. and Santos, Jorge E.",
    title = "{Sudden breakdown of effective field theory near cool Kerr-Newman black holes}",
    eprint = "2403.00051",
    archivePrefix = "arXiv",
    primaryClass = "hep-th",
    doi = "10.1007/JHEP05(2024)122",
    journal = "JHEP",
    volume = "05",
    pages = "122",
    year = "2024"
}

@article{Reall:2019sah,
    author = "Reall, Harvey S. and Santos, Jorge E.",
    title = "{Higher derivative corrections to Kerr black hole thermodynamics}",
    eprint = "1901.11535",
    archivePrefix = "arXiv",
    primaryClass = "hep-th",
    doi = "10.1007/JHEP04(2019)021",
    journal = "JHEP",
    volume = "04",
    pages = "021",
    year = "2019"
}

@article{Goon:2016mil,
    author = "Goon, Garrett",
    title = "{Heavy Fields and Gravity}",
    eprint = "1611.02705",
    archivePrefix = "arXiv",
    primaryClass = "hep-th",
    doi = "10.1007/JHEP01(2017)045",
    journal = "JHEP",
    volume = "01",
    pages = "045",
    year = "2017",
    note = "[Erratum: JHEP 03, 161 (2017)]"
}

@article{Bellazzini:2015cra,
    author = "Bellazzini, Brando and Cheung, Clifford and Remmen, Grant N.",
    title = "{Quantum Gravity Constraints from Unitarity and Analyticity}",
    eprint = "1509.00851",
    archivePrefix = "arXiv",
    primaryClass = "hep-th",
    reportNumber = "CALT-TH-2015-044, SACLAY-T15-161",
    doi = "10.1103/PhysRevD.93.064076",
    journal = "Phys. Rev. D",
    volume = "93",
    number = "6",
    pages = "064076",
    year = "2016"
}

@article{Gruzinov:2006ie,
    author = "Gruzinov, A. and Kleban, M.",
    title = "{Causality Constrains Higher Curvature Corrections to Gravity}",
    eprint = "hep-th/0612015",
    archivePrefix = "arXiv",
    doi = "10.1088/0264-9381/24/13/N02",
    journal = "Class. Quant. Grav.",
    volume = "24",
    pages = "3521--3524",
    year = "2007"
}

@article{MacGibbon:1987my,
    author = "MacGibbon, Jane H.",
    title = "{Can Planck-mass relics of evaporating black holes close the universe?}",
    doi = "10.1038/329308a0",
    journal = "Nature",
    volume = "329",
    pages = "308--309",
    year = "1987"
}

@article{Endlich:2017tqa,
    author = "Endlich, Solomon and Gorbenko, Victor and Huang, Junwu and Senatore, Leonardo",
    title = "{An effective formalism for testing extensions to General Relativity with gravitational waves}",
    eprint = "1704.01590",
    archivePrefix = "arXiv",
    primaryClass = "gr-qc",
    doi = "10.1007/JHEP09(2017)122",
    journal = "JHEP",
    volume = "09",
    pages = "122",
    year = "2017"
}

@article{Hawking:1975vcx,
    author = "Hawking, S. W.",
    editor = "Gibbons, G. W. and Hawking, S. W.",
    title = "{Particle Creation by Black Holes}",
    doi = "10.1007/BF02345020",
    journal = "Commun. Math. Phys.",
    volume = "43",
    pages = "199--220",
    year = "1975",
    note = "[Erratum: Commun.Math.Phys. 46, 206 (1976)]"
}

@article{Horowitz:2022mly,
    author = "Horowitz, Gary T. and Kolanowski, Maciej and Santos, Jorge E.",
    title = "{Almost all extremal black holes in AdS are singular}",
    eprint = "2210.02473",
    archivePrefix = "arXiv",
    primaryClass = "hep-th",
    doi = "10.1007/JHEP01(2023)162",
    journal = "JHEP",
    volume = "01",
    pages = "162",
    year = "2023"
}

@article{Nomura:2012cx,
    author = "Nomura, Yasunori and Varela, Jaime and Weinberg, Sean J.",
    title = "{Black Holes, Information, and Hilbert Space for Quantum Gravity}",
    eprint = "1210.6348",
    archivePrefix = "arXiv",
    primaryClass = "hep-th",
    reportNumber = "MIT-CTP-4405, UCB-PTH-12-17",
    doi = "10.1103/PhysRevD.87.084050",
    journal = "Phys. Rev. D",
    volume = "87",
    pages = "084050",
    year = "2013"
}

@article{Barausse:2018vdb,
    author = "Barausse, Enrico and Brito, Richard and Cardoso, Vitor and Dvorkin, Irina and Pani, Paolo",
    title = "{The stochastic gravitational-wave background in the absence of horizons}",
    eprint = "1805.08229",
    archivePrefix = "arXiv",
    primaryClass = "gr-qc",
    doi = "10.1088/1361-6382/aae1de",
    journal = "Class. Quant. Grav.",
    volume = "35",
    number = "20",
    pages = "20LT01",
    year = "2018"
}

@article{Page:1976df,
    author = "Page, Don N.",
    title = "{Particle Emission Rates from a Black Hole: Massless Particles from an Uncharged, Nonrotating Hole}",
    doi = "10.1103/PhysRevD.13.198",
    journal = "Phys. Rev. D",
    volume = "13",
    pages = "198--206",
    year = "1976"
}

@article{Xin2026,
  title = {Relativistic tidal tensor: General solutions for stationary axisymmetric spacetimes and the Hills mass of naked singularities},
  author = {Xin, Wenkang and Mummery, Andrew},
  journal = {Phys. Rev. D},
  volume = {113},
  issue = {4},
  pages = {044036},
  numpages = {19},
  year = {2026},
  month = {Feb},
  publisher = {American Physical Society},
  doi = {10.1103/xdm3-48nb},
  url = {https://link.aps.org/doi/10.1103/xdm3-48nb}
}

@article{Hills1975,
  author = {Hills, J. G.},
  title = {Possible power source of Seyfert galaxies and QSOs},
  journal = {Nature},
  year = {1975},
  volume = {254},
  pages = {295-298},
  doi = {10.1038/254295a0}
}

\end{document}